\documentclass[review]{elsarticle}

\usepackage{lineno,hyperref}
\modulolinenumbers[5]

\usepackage{xcolor}
\usepackage{amsmath,amssymb}

\journal{-}

\begin{document}

\begin{frontmatter}

\title{The role of polymer structure on water confinement in poly(\emph{N}-isopropylacrylamide) dispersions\tnoteref{mytitlenote}}

\author[address1]{E. Buratti}

\author[address1]{L. Tavagnacco}

\author[address2]{M. Zanatta}

\author[address3]{E. Chiessi}

\author[address4]{S. Buoso}

\author[address1]{S. Franco}

\author[address1]{B. Ruzicka}

\author[address1]{R. Angelini}

\author[address5]{A. Orecchini}

\author[address6]{M. Bertoldo}

\author[address1]{E. Zaccarelli\corref{mycorrespondingauthor}}
\cortext[mycorrespondingauthor]{Corresponding author}
\ead{emanuela.zaccarelli@cnr.it}

\address[address1]{CNR-ISC and Department of Physics, Sapienza University of Rome, I-00185 Roma, Italy}
\address[address2]{Department of Physics, University of Trento, I-38123 Trento, Italy}
\address[address3]{Department of Chemical Sciences and Technologies, University of Rome Tor Vergata, I-00133 Roma, Italy}
\address[address4]{ISOF-CNR Area della Ricerca di Bologna, I-40129 Bologna, Italy}
\address[address5]{Department of Physics and Geology, University of Perugia, I-06123, Perugia, Italy}
\address[address6]{Department of Chemical, Pharmaceutical and Agricultural Sciences, University of Ferrara, I-44121 Ferrara, Italy}

\begin{abstract}

Poly(\emph{N}-isopropylacrylamide) (PNIPAM) is a synthetic polymer that is widely studied for its thermoresponsive character. However, recent works also reported evidence of a low temperature (protein-like) dynamical transition around 225 K in concentrated PNIPAM suspensions, independently of the polymer architecture, i.e., both for linear chains and for microgels. In this work, we investigate water-polymer interactions by extensive differential scanning calorimetry (DSC) measurements of both systems, in order to understand the effect of the different topological structures on the solution behaviour, in particular regarding crystallization and melting processes. In addition, we compare protiated and
deuterated microgels, in both water and deuterated water. The DSC results are complemented by dynamic light scattering experiments, which confirm that the selective isotopic substitution differently affects the solution behaviour. Our findings highlight the important role played by the polymer architecture on the solution behaviour: indeed, microgels turn out to be more efficient confining agents, able to avoid water crystallization in a wider concentration range with respect to linear chains. Altogether, the present data will be valuable to interpret future low-temperature investigations of PNIPAM dispersions, particularly by neutron scattering experiments.
\end{abstract}


\end{frontmatter}

\section{Introduction}

Poly(\emph{N}-isopropylacrylamide) (PNIPAM) is a thermoresponsive polymer whose aqueous solutions show a Lower Critical Solution Temperature (LCST) around 305 K~\cite{halperin2015poly}, where the polymer undergoes a conformational transition: below it, chains are hydrated and dispersed in water; above it, they become insoluble, leading to a sharp but reversible coil-to-globule transition.

Several works have highlighted that PNIPAM, due to its amphiphilic character and the chemical analogy to polypeptides, can be used as model to mimic protein folding and denaturation~\cite{tiktopulo1995domain,fujishige1989phase,graziano2000temperature,rosi2021thermoresponsivity}. In particular, a recent investigation by means of elastic incoherent neutron scattering (EINS) measurements on PNIPAM chains at different degree of hydration reported evidence of the occurrence of a protein-like dynamical transition around 225 K~\cite{tavagnacco2021proteinlike}. This transition was observed upon cooling PNIPAM dispersions in D$_2$O down to very low temperatures, where notably solvent crystallization was avoided at high enough polymer concentration.
Related works investigated the role of polymeric architecture on this transition, which has been detected both in linear chains and microgels~\cite{zanatta2018evidence,tavagnacco2019water,tavagnacco2021proteinlike}.
Microgels are colloidal particles consisting of a cross-linked polymer network, with a structure that is typically characterized by a dense core surrounded by a loose corona~\cite{BookFernandez2011,stieger2004small}. In microgels the analogue of LCST demixing is represented by the occurrence of a so-called Volume Phase Transition (VPT) from a swollen to a collapsed state~\cite{pelton2010poly}, which is reversible and allows to tune the size of the particles, a very appealing feature for both fundamental science~\cite{yunker2014physics,rovigatti2019numerical,brijitta2019responsive} and a variety of nano and bio-technological applications~\cite{oh2008development,fernandez2009gels,karg2019nanogels,stuart2010emerging}.
Moreover, microgels internal architecture appears to be particularly suitable to efficiently confine water molecules~\cite{afroze2000phase,van2004kinetics,zanatta2018evidence}  and avoid their crystallization down to very low temperatures. This is important both for fundamental science, as it could provide a way to investigate water properties in the so-called no man's land~\cite{mishima1998relationship}, where interesting phenomena, such as a liquid-liquid phase transition~\cite{debenedetti2020second}, should take place and of which there is so far no direct evidence. In addition, the understanding of the mechanism that leads to avoided water crystallization is also relevant in biological systems~\cite{knight2000adding} for a variety of applications, ranging for example from food industry~\cite{kiani2011water,kang2020supercooling} to tissues and organs preservation~\cite{tas2021freezer}. To this aim, it appears extremely interesting to investigate the high-concentration and low-temperature regime of microgel suspensions, in order to better understand their interplay with solvent crystallization. Notwithstanding the wide literature on PNIPAM, this regime is still rather unexplored.

Differential Scanning Calorimetry (DSC) can be a very useful technique to probe the evolution of the hydration properties of these systems and to monitor the behaviour of bulk and hydration water. However, DSC was mostly applied so far to study the temperature-induced phase separation of PNIPAM aqueous solutions~ \cite{fujishige1989phase,tiktopulo1994cooperativity, schild1990microcalorimetric, kujawa2001volumetric, ding2006microcalorimetric, qiu2013new, sun2013effect, bischofberger2014hydrophobic}, typically at very low polymer concentrations and in limited temperature intervals across the LCST. Only a scarce number of papers reported on the phase behaviour, as detected by DSC, of PNIPAM linear chains in water~ \cite{van2004kinetics} and of chemically cross-linked PNIPAM hydrogels~ \cite{afroze2000phase} in a wider range of polymer concentration, extending the analysis to temperatures far below the water crystallization conditions.

In this work, we investigate the role of polymer structure, water content and water-polymer interactions on the VPT as well as on the melting and crystallization processes. To this aim, we perform extensive DSC measurements of PNIPAM suspensions in a wide range of polymer concentration, from 1\% up to 80\% by weight and in a temperature interval between 193 K and 333 K. Moreover, we study dispersions of both PNIPAM linear chains and PNIPAM microgels, to understand the effect of the architecture on the solution behaviour and on the capability of efficiently confining water. In order to be able to compare our results with those obtained from neutron scattering experiments, polymer chains and microgels were dispersed both in water and in deuterium oxide. Since we work at rather high polymer concentrations, we carefully study the aging of the samples, repeating the DSC measurements until reaching a stationary state. Also, to further investigate water-polymer interactions, properties of protiated and deuterated microgels were compared. Finally, DSC results are complemented by dynamic light scattering (DLS) experiments in order to properly characterize the VPT of the different samples.
Our results show that both the polymer structure and the isotope substitution have an effect on the interactions with the solvent molecules and influence the upper limit of polymer concentration at which water is able to crystallize. We observe some differences also in the low-concentration regime, especially regarding the PNIPAM conformational transition, where the deuterium substitution plays a very relevant role.

\section{Materials and methods}

\emph{N}-isopropylacrylamide (NIPAM) (Sigma-Aldrich, St. Louis, MO, USA), purity 97\%, and deuterated \emph{N}-isopropylacrylamide (d-10) (d-NIPAM) (Polymer Source, Inc, Quebec, Canada), purity $\geq 98\%$, were purified by recrystallization from hexane, dried under reduced pressure (0.01 mmHg) at room temperature and stored at 253 K. N,N’-methylenebisacrylamide (BIS) (Sigma-Aldrich, St. Louis, MO, USA), electrophoresis grade, was purified by recrystallization from methanol, dried under reduced pressure (0.01 mmHg) at room temperature and stored at 253 K. Sodium dodecyl sulphate (SDS), purity 98\% and potassium persulfate (KPS), purity 98\% were purchased from Sigma-Aldrich (St. Louis, MO, USA) and used as received.
Poly(\emph{N}-isopropylacrylamide) linear chains Mw=189 kDa (P$_{lin189}$) (PDI 2.88)  (Polymer Source, Inc, Quebec, Canada) and Mn=40 kDa (P$_{lin40}$) (Sigma-Aldrich, St. Louis, MO, USA) were used as received.
Ultrapure water (resistivity: 18.2 MW/cm at room temperature) was obtained with Arium\textsuperscript{®} pro Ultrapure water purification Systems, Sartorius Stedim.
All the other solvents (Sigma Aldrich RP grade) were used as received. Dialysis membrane, SpectraPor\textsuperscript{®} 1, MWCO 6-8 kDa (Spectrum Laboratories, Inc., Piscataway, NJ, USA) was soaked in distilled water for 2 h and then thoroughly rinsed before use.

\subsection{Microgels synthesis}

\begin{figure}[hbtp]
\centering
\includegraphics[scale=0.5]{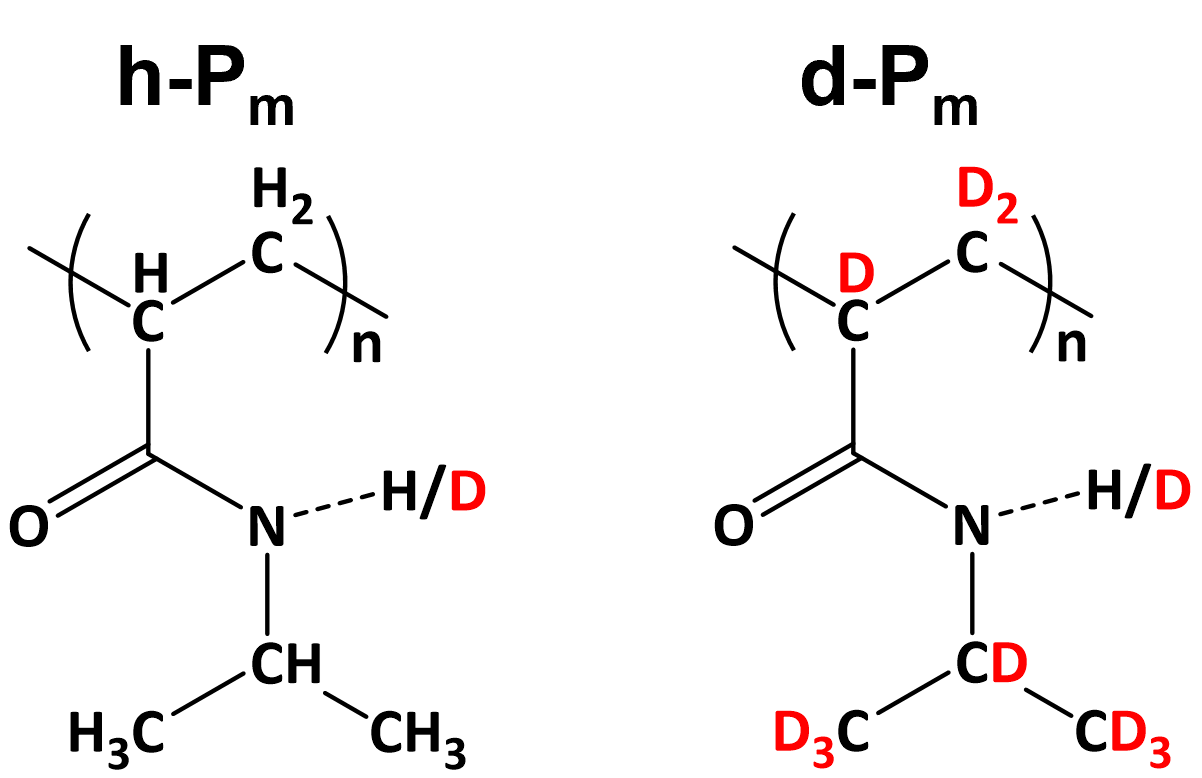}
\caption{Chemical structures of the main repeating unit in the protiated (on the left) and the deuterated (on the right) PNIPAM microgels.}
\label{structures}
\end{figure}

Two different PNIPAM microgels, a protiated and a deuterated one, were synthesised by standard precipitation polymerization. The chemical structures of the main repeting units is reported in Fig.~\ref{structures}. In the case of the non-deuterated microgel, that we refer to as h-P$_m$, NIPAM (0.137 M), BIS (1.87 mM) and SDS (7.82 mM) were solubilized in 1560 mL of ultrapure water into a 2 L jacket reactor. The solution was deoxygenated by bubbling nitrogen for 1 h and then heated at 343 K. Initiator KPS (2.44 mM) was dissolved in 10 mL deoxygenated water and added to initiate the polymerization, that was left to proceed for 4 h. Deuterated PNIPAM microgel, labelled as d-P$_m$, was synthesised using the same molar concentration of reactants but substituting NIPAM with NIPAM-d10. As the crosslinker concentration is much smaller than the monomer one, non-deuterated BIS was used for both microgels. The crude dispersions were purified by dialysis (MWCO 6–8 kDa) with distilled water with frequent water change for 2 weeks, then they were concentrated by lyophilisation up to 10 wt\% in H$_2$O. To prepare samples in deuterium oxide, two cycles of lyophilisation to dryness and redispersion in D$_2$O (10 wt\%) were performed, in order to ensure the quantitative removal of H$_2$O.

\subsection{Differential Scanning Calorimetry}

Thermal analyses on PNIPAM samples, both crosslinked microgels and linear polymers, were performed with a Perkin Elmer Pyris Diamond DSC equipped with Intracooler III as cooling system and a DSC 8000 Perkin Elmer differential scanning calorimeter equipped with Intracooler II as cooling system. About 10-15 mg of PNIPAM dispersions at different concentrations was analysed under nitrogen atmosphere (20 mL/min) in hermetic sealed steel pans, to prevent changes in concentration during the heating/cooling steps. The weight of sample containing pans was checked after experiments and no changing in weight was observed. Samples were prepared by pouring a dispersion at 10 wt\% (either in H$_2$O or D$_2$O) into the pans and then evaporating the exceeding solvent in a desiccator, until the desired concentration was reached. The concentration was regularly monitored by accurate weighing of the samples. Once reached the target concentration, pans were hermetically sealed and analysed.
The measurements were carried out by cooling the system from 293 to 193 K, holding the temperature for 3 minutes, then heating it to 333 K, holding the temperature for 3 minutes and finally by cooling it again to 293 K. Each cooling and heating step was carried out with a scanning rate of 5 K/min. Additional scanning rates of 10 and 20 K/min were employed for assessing the onset of the glass transition.
Several cycles of cooling and heating were performed on each sample in order to investigate the aging behaviour of all detectable transitions and whether a long-time stationary state can be reached. The details of the investigation are reported in the Supplementary Material. In the remaining of this article, we only focus on data taken after several thermal cycles and in long-time measurements at all concentrations, thus avoiding the interference of aging effects. In addition, the procedure that we used to analyse all the transitions in the thermograms, in particular regarding the melting peak, is also described in more detail in the Supplementary Material.

\subsection{Dynamic Light Scattering}

Dynamic Light Scattering (DLS) measurements were performed on non-deuterated and deuterated PNIPAM microgels at concentration of 0.01\% by weight in both H$_2$O and in D$_2$O, in a range of temperatures between 293 K and 317 K.
Measurements were carried out using an optical setup based on a solid state laser (100 mW) with a monochromatic ($\lambda$ = 642 nm) and polarized beam, at a scattering angle $\theta$ = 90° corresponding to a scattering vector $Q = 0.018 \, \mathrm{nm}^{-1}$, according to the relation $Q = (4 \pi n / \lambda) \sin (\theta / 2)$. The hydrodynamic radius of the particles was obtained through the Stokes-Einstein relation $R = k_B T/6\pi \eta D_t$, where $k_B$ is the Boltzmann constant, $\eta$ the viscosity and $D_t$ the translational diffusion coefficient, related to the relaxation time $\tau$ through the relation $\tau = 1/(Q^{2} D_t)$. The relaxation time $\tau$ was obtained by fitting the autocorrelation function of scattered intensity through the Kohlrausch-William-Watts expression, $g_2(Q,t) = 1 + b[\exp(-(t/\tau)^{\beta})]^2$, as commonly used for colloidal systems with low polydispersity~\cite{nigro2019study}, where $\beta$ is the stretching exponent.

\section{Results}

\begin{figure}[hbtp]
\centering
\includegraphics[width=8.5cm]{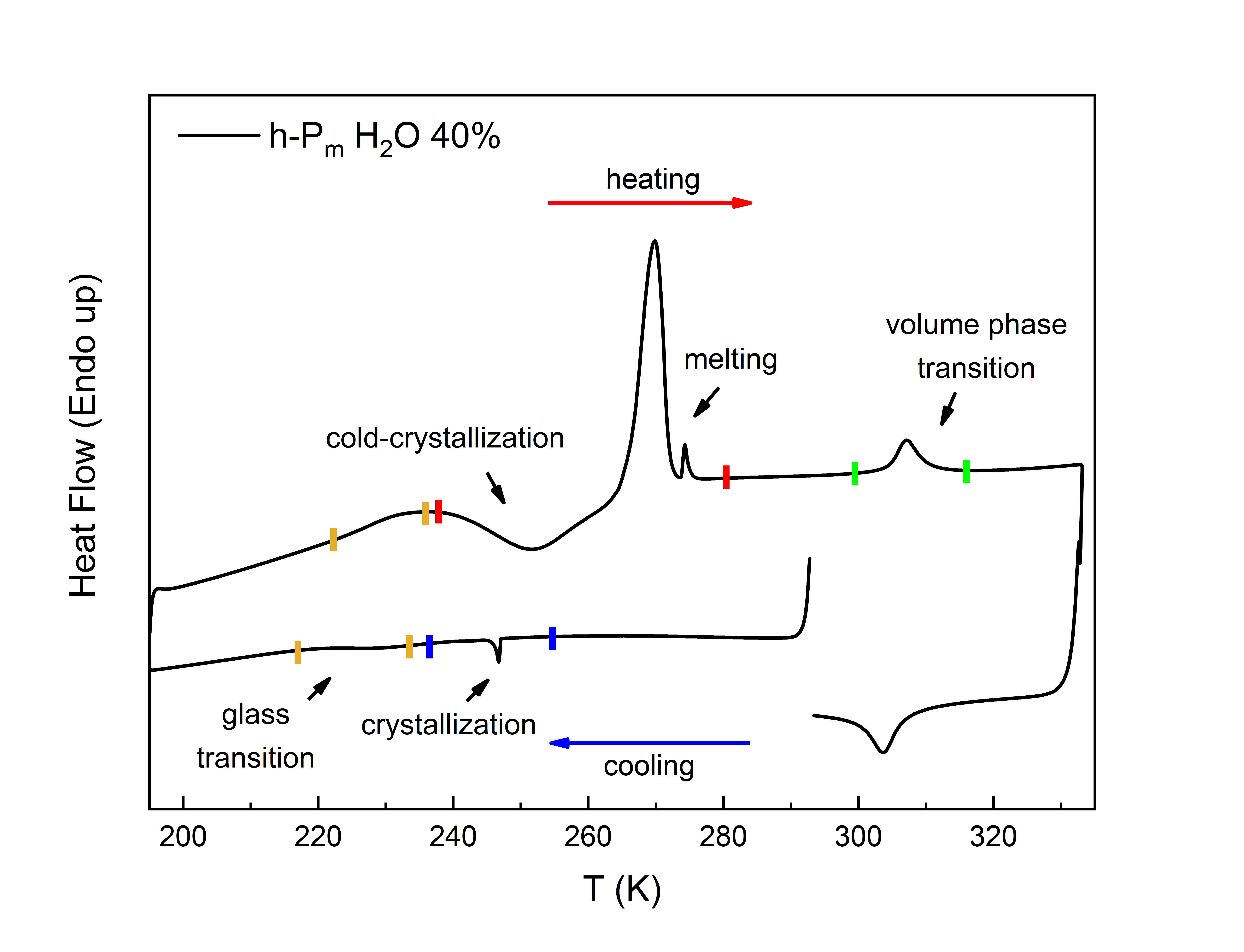}
\caption{Thermogram of h-P$_m$ microgels at a concentration of 40 wt\% in H$_2$O.}
\label{thermogram}
\end{figure}

We start by reporting a representative DSC thermogram of h-P$_m$ microgels at 40\% weight in Fig.~\ref{thermogram}. It can be observed that, upon cooling the dispersion from room temperature down to very low temperature (193 K), crystallization of water occurs, as evident from the exothermic peak between the blue lines. At lower temperatures, a glass transition appears, manifesting itself as a change of slope (in the thermogram between orange lines), both under cooling and then under heating. Then upon re-heating up to room temperature, the melting (endothermic peaks between red lines) of the previously crystallized solvent is noticed, preceded by a process of cold-crystallization (exothermic peak).
A further increase of temperature leads to the PNIPAM conformational transition around 305 K, associated to the VPT in this case, that is shown by the endothermic peak between green lines.

\subsection{Study of the water melting}

In previous works~\cite{zanatta2018evidence,tavagnacco2021proteinlike}, we observed the occurrence of a protein-like dynamical transition in PNIPAM microgels as well as in PNIPAM linear chains at high concentrations, above roughly 40 wt\%, where the crystallization of water was absent at low temperatures. These results are in  qualitative agreement with previous DSC measurements~\cite{afroze2000phase}, but we aim to be a bit more quantitative on such crystallization in order to highlight the role of the polymer architecture on water confinement and on the suppression of crystallization.

\begin{figure}[hbtp]
\centering
\includegraphics[width=8.5cm]{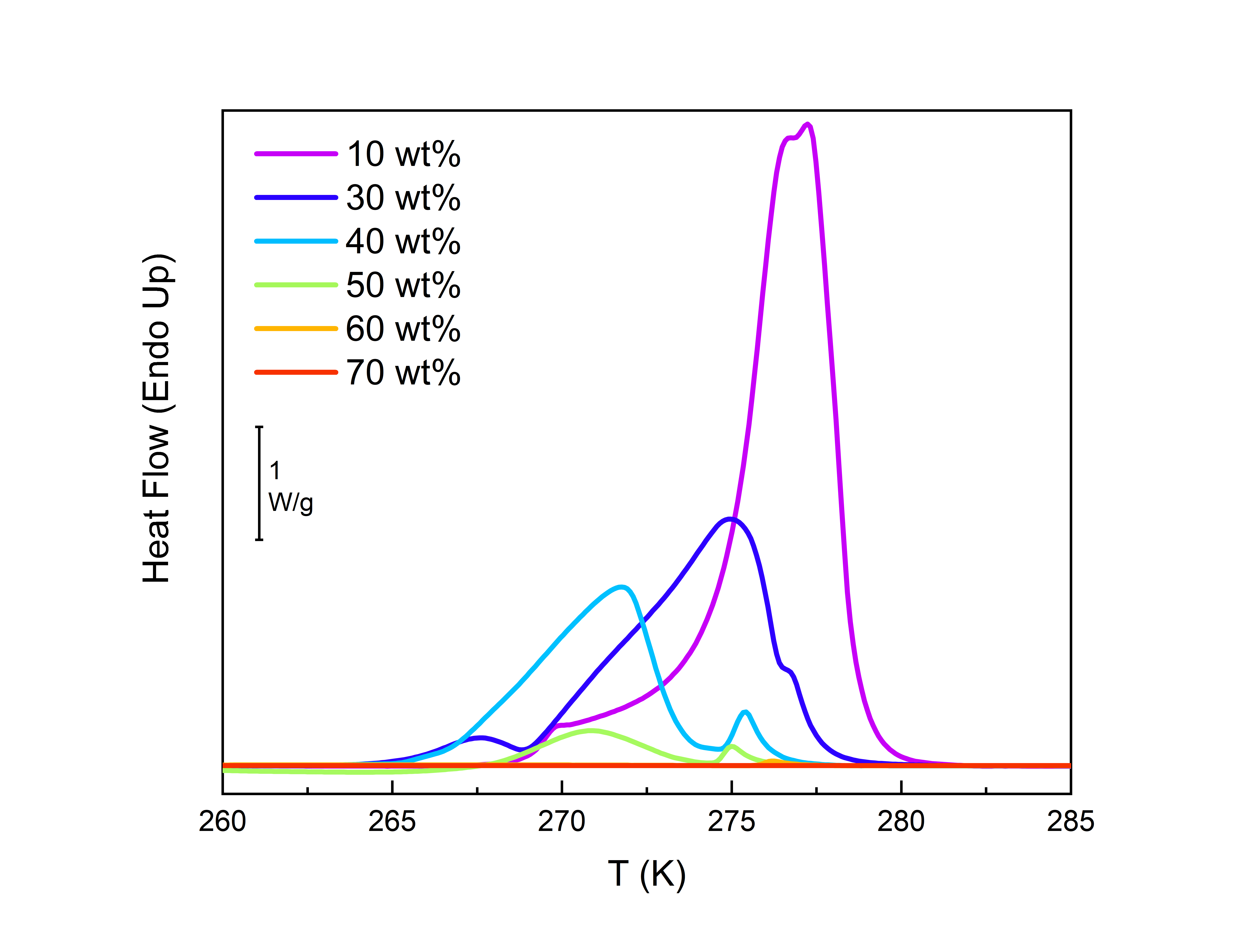}
\caption{Melting peaks of h-P$_m$ in D$_2$O for samples in the range of concentration 10 – 70 wt\%. Thermograms are normalized to the mass of dispersion.}
\label{melting}
\end{figure}

Therefore, we now provide a careful investigation of this issue, aiming to establish in which range of polymer concentration water crystallization is suppressed. By means of calorimetric measurements, we can specifically obtain quantitative information on the solvent behaviour through the analysis of its melting peak. In Fig.~\ref{melting}, as an example, we report the evolution of the melting peak as a function of polymer concentration for the non-deuterated microgels in D$_2$O.
It can be seen that there is a shift of the main peak toward lower temperatures by increasing concentration, associated to a decrease of the peak area, in agreement with the lower amount of solvent in more concentrated dispersions. In addition,
for PNIPAM concentrations higher than 30 wt\%, two endothermic peaks are observed: one in the range 268 – 274 K and the other around 276 K. The latter peak is always found approximately at this temperature value for samples in D$_2$O, while it is located around 274 K for samples in H$_2$O.

\begin{figure}[hbtp]
\centering
\includegraphics[width=\textwidth]{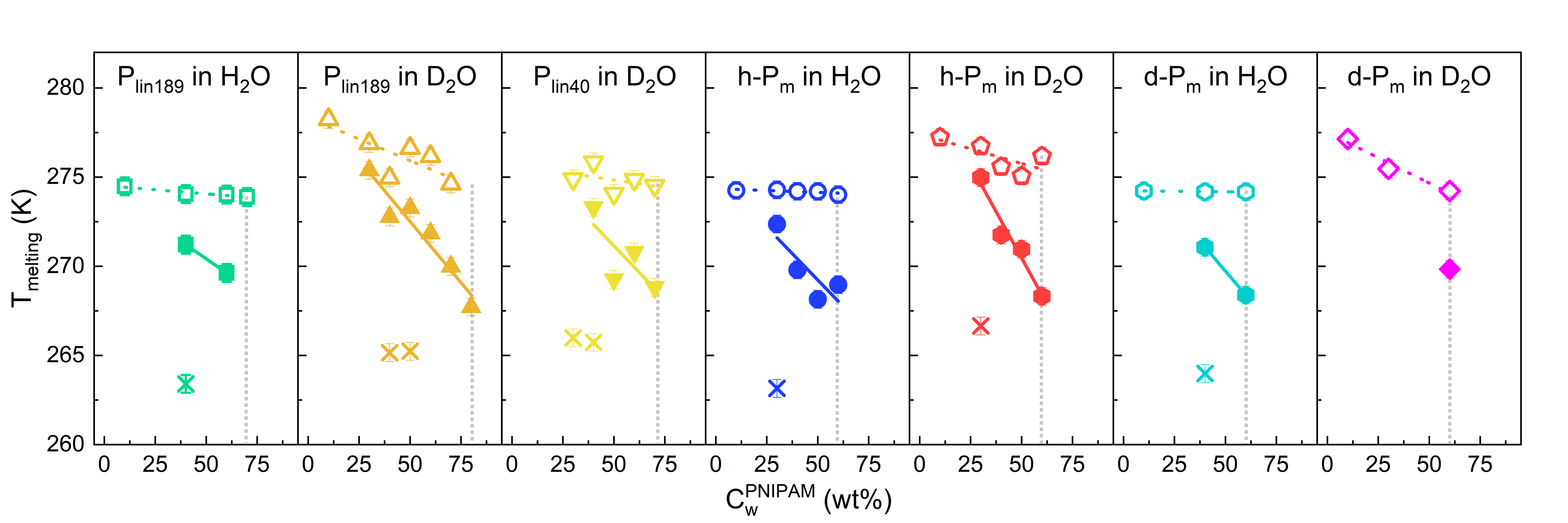}
\caption{Melting temperature of the ``free water'' (open symbols) and of the ``freezable-bound water'' (closed symbols) for all studied polymer-solvent combinations as a function of PNIPAM concentration. Lines are guides to the eye. The melting temperature of the third peak appearing only at intermediate concentrations is represented by crosses. The vertical dotted lines represent the lower limit above which solvent crystallization is completely absent. Errors on temperature values are estimated to be of the order of 0.5 K. Error bars are reported with the data but they are not clearly visible in the scale of the figure.}
\label{Tmelting}
\end{figure}

To better visualize the evolution of the $T$-dependence of these melting peaks, we plot in Fig.~\ref{Tmelting} the temperatures associated to the maximum of the calorimetric peaks as a function of polymer concentration for all investigated structures and dispersions.
All studied samples point to the fact that there is evidence of melting of two different types of water, in agreement with previous works~\cite{ping2001states}.
Indeed, the peak at higher temperature (open symbols) can be associated to the so-called ``free water'' (bulk-like water), since it has a melting point roughly equal to the one of the pure solvent ($\sim 274$ K for H$_2$O and $\sim 276$ K for D$_2$O) and it is found to be quite independent of concentration.
On the other hand, the peak observed at lower temperature (closed symbols) is much more affected by the presence of the polymer and its melting temperature shifts toward lower values by increasing PNIPAM concentration. For these reasons, it appears to be associated to crystallizing hydration water, which has previously been referred as ``freezable-bound water''~\cite{ping2001states}.
The onset of these two types of water is general and happens for all the studied samples, regardless of the polymer architecture and type of solvent.
We further notice in Fig.~\ref{melting} that, only for concentrations around 30-40 wt\%, a third, additional peak also arises, of very small intensity, at even lower temperatures. This peak is present in this range of concentrations for all different samples, thus appearing a robust feature of the thermograms. This could be tentatively associated to yet another type of water affected by the polymer interaction~\cite{hatakeyma2007cold}, for example located in a further proximal domain, possible only at intermediate concentrations.
At higher concentrations, the peak disappears since in that case it is not possible to coordinate water molecules in the more distant coordination shells. A second hypothesis is that this peak may be due to small crystals of confined water. This might explain the behaviour of the melting temperature, which is well known to decrease with crystal size, due to the predominant effect of the excess of energy at the crystal surface.

While the temperature of the melting peaks is relatively unaffected by the polymer structure, we find a clear indication that this plays an important role in the value of the critical concentration above which melting was no longer detected, as also reported in Fig.~\ref{Tmelting}.
We note that such a critical value is significantly lower for PNIPAM microgels (around 60 wt\%) than for linear chains (70-80 wt\%), suggesting that the topological constraint induced by the presence of crosslinks in the network acts against proper water-water hydrogen bonds rearrangements and thus crystallization. In the case of P$_{lin189}$ and P$_{lin40}$, instead, the water molecules and the linear chains are more free to move and rearrange and thus the critical value is shifted to higher concentrations. Interestingly, we do not observe significant differences in the concentration threshold for chains at the two studied molecular weights, in agreement with previous studies~\cite{afroze2000phase}. Similarly, no significant differences in the value of the cystallization limit are observed on changing solvent from water to deuterium oxide and/or using deuterated microgels.

\begin{figure}[htp]
\centering
\includegraphics[width=8.5cm]{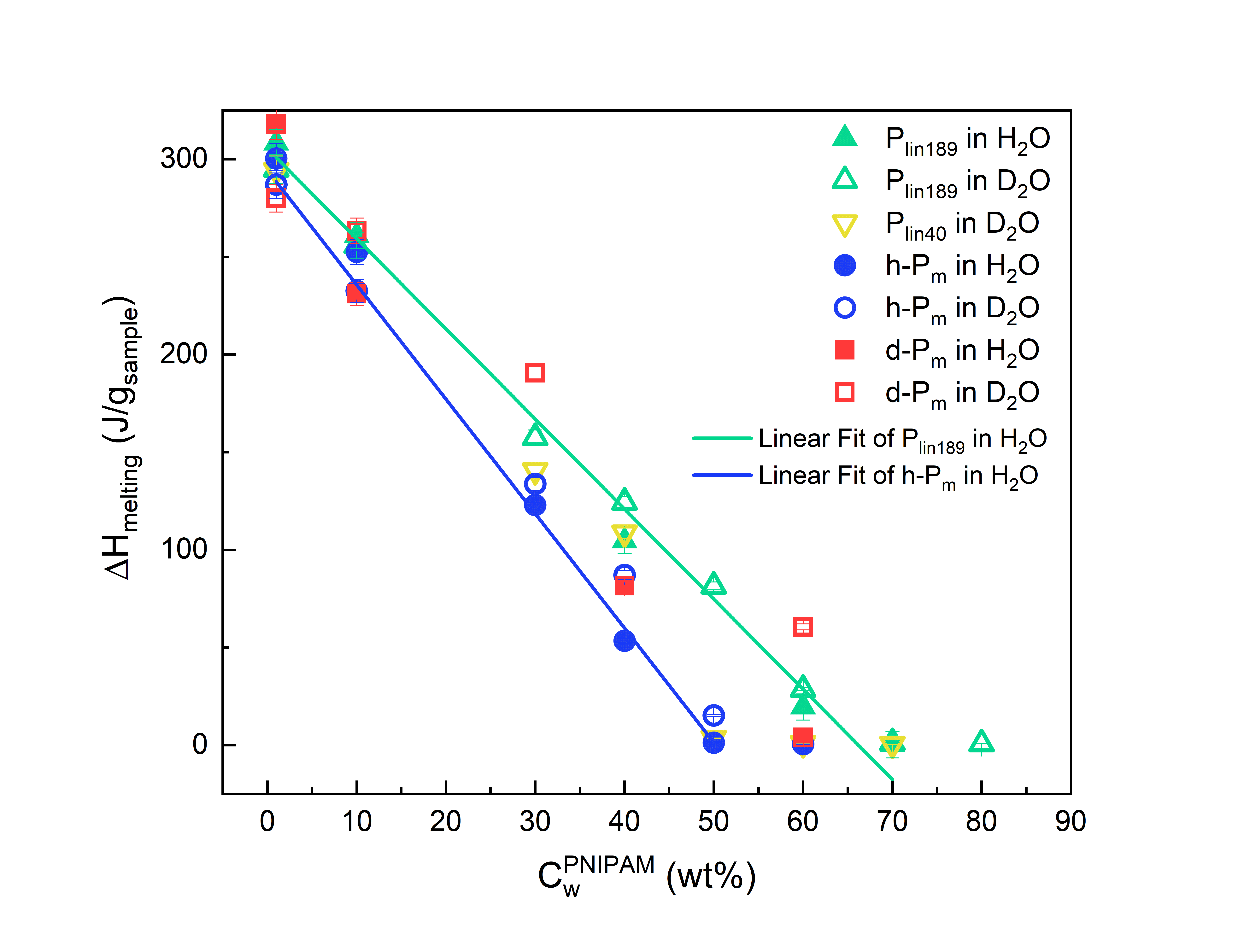}
\caption{Melting enthalpy of the total peak as a function of polymer concentration for all studied samples. Green and blue lines are the linear regression of the data of P$_{lin189}$ in H$_2$O and h-P$_m$ in H$_2$O, respectively. Estimated slopes are -4.6 $\pm$ 0.3 and -5.9 $\pm$ 0.3, respectively. The percentage error on the enthalpy values is estimated to be around 2.5\%. Error bars are reported with the data but they are not clearly visible in the scale of the figure.}
\label{meltingenthalpy}
\end{figure}

We now turn to examine the behaviour of the peak area and of the associated melting enthalpy, which is found to decrease with polymer concentration, as expected. In Fig.~\ref{meltingenthalpy}, the melting enthalpy of the entire peak is reported for the dispersions at different concentrations. For the most dilute conditions, enthalpy is around 300 J/g, close to the value of the pure solvents, as also previously confirmed in Ref.~\cite{afroze2000phase}. Then a roughly linear decrease is observed, that appears to be steeper for microgels than for linear chains, as we can see in the figure from the comparison of the linear fits of P$_{lin189}$ and h-P$_m$ both in H$_2$O, until it reaches zero at polymer concentration around 60 or 75 wt\%, respectively.

In order to further analyse the interactions of the microgels and of the linear chains with the ``free'' and ``freezable-bound'' water , we developed a procedure to separate the areas of the corresponding peaks, as described in more detail in the Supplementary Material.

\begin{figure}[ht]
\centering
\includegraphics[width=\textwidth]{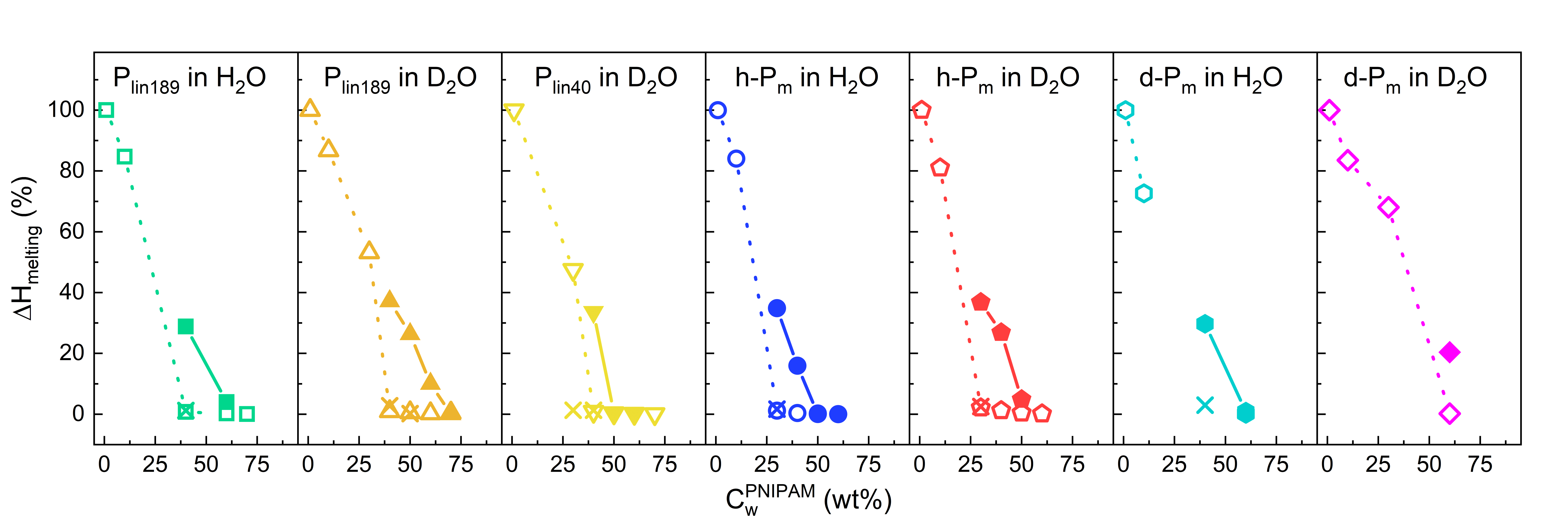}
\caption{Normalized melting enthalpy of free water (open symbols) and freezable-bound water (closed symbols) for each investigated polymer-solvent case. Lines are guides to the eye. The melting enthalpy associated to the third peak occurring at intermediate concentrations is represented by the crosses. The area of the peak of 1\% samples is taken as the reference case at 100\%.}
\label{Hmelting}
\end{figure}

The resulting fraction of melting enthalpy associated to each peak is reported in Fig.~\ref{Hmelting}. Data are normalized to the results at the most dilute investigated concentration (1 wt\%).
We observe that free water (open symbols) is the only type of water present at low concentration, while the onset of freezable-bound water arises for all samples around 30-40 wt\%. Above this value, the enthalpy associated to free water rapidly drops while that related to freezing hydration water remains detectable up to the high concentration limit of crystallization.

While we have examined the tendency to crystallize, we have not been quantitative on this aspect yet. The degree of crystallinity $\chi_c$ in the dispersion can be estimated through the following equation \cite{kong2002measurement}:
\begin{equation}\label{Eq:chi}
\chi_c \, (\mathrm{wt\%}) = \frac{\Delta H_m}{\Delta H^{\circ}_m} \frac{1}{\left(1-\frac{C_w^{\mathrm{PNIPAM}}}{100} \right)} 100 ,
\end{equation}
where $\Delta{H^{\circ}_m}$ is the standard fusion enthalpy either of bulk H$_2$O at 273 K, that is 333 J/g, or of bulk D$_2$O at 276.7 K, that is 340.7 J/g, (1 - $\frac{C^{\mathrm{PNIPAM}}_w}{100}$) is the weight fraction of H$_2$O or D$_2$O in the sample and $\Delta{H_m}$ (J/g$_{\mathrm{sample}}$) is the enthalpy associated with the overall peak, that is the sum of ``free water'' and ``freezable-bound water''.
It is important to note that the values of $\chi_c$ obtained in this way are entirely attributable to the solvent, since poly(\emph{N}-isopropylacrylamide) is an amorphous polymer, and hence by this quantity we can directly compare results for different polymer structures and solvent content.

\begin{figure}[hbtp]
\centering
\includegraphics[width=8.5cm]{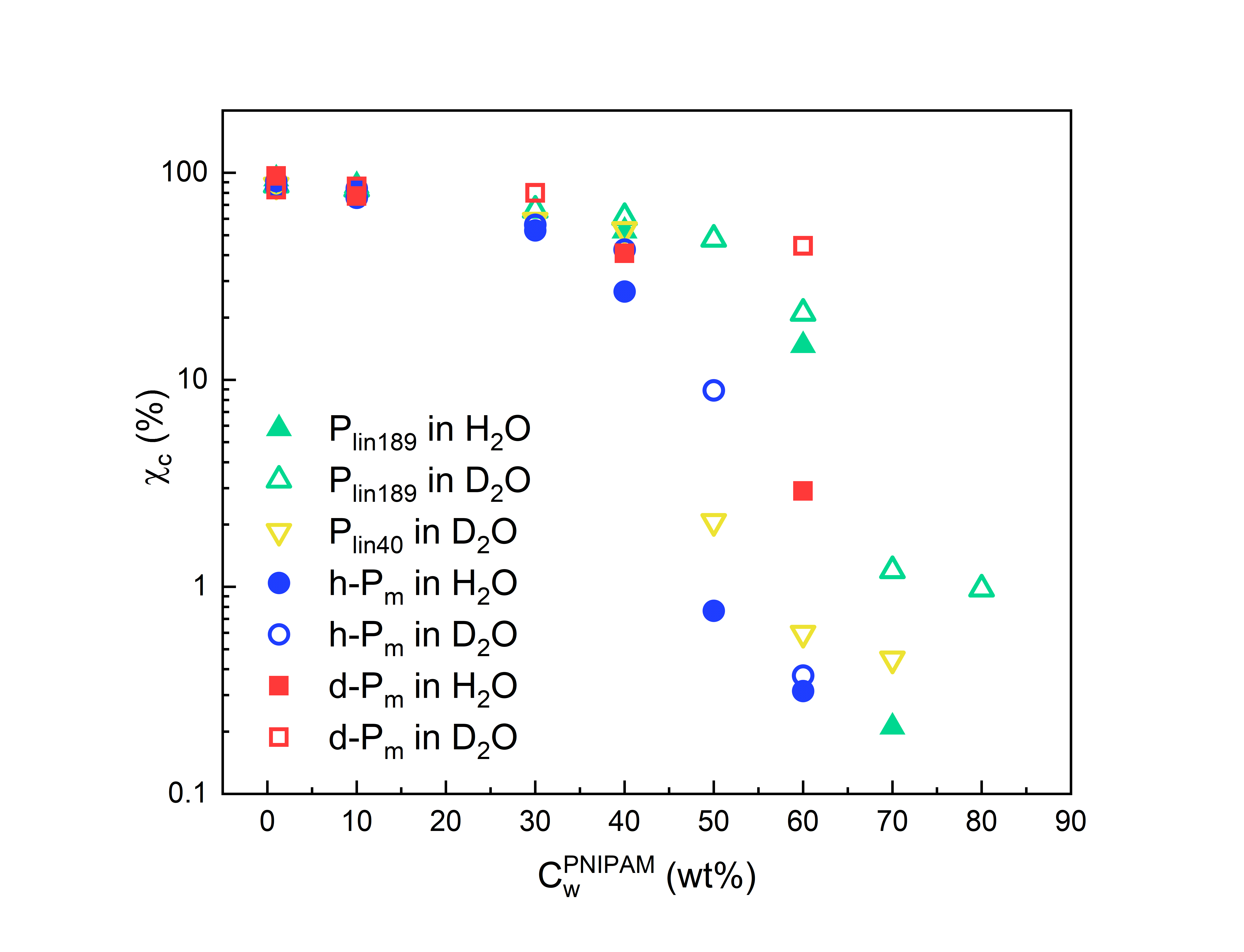}
\caption{Crystallization degree, $\chi_c$, as a function of polymer concentration for each sample, calculated from Equation~\ref{Eq:chi}.}
\label{crystallinity}
\end{figure}

Fig.~\ref{crystallinity} shows that similar crystallinity degrees are found at high solvent content for all the samples, while clear differences are present upon increasing PNIPAM concentration above 40 wt\%. In agreement with the information obtained from the melting temperature, we find that $\chi_c$ is significantly larger for linear chains than for microgels, indicating that crystallization is facilitated by the higher mobility of the chains.  Results for the two different molecular weights of linear PNIPAM are similar to each other, in agreement with previous work~\cite{afroze2000phase}. However, a slight difference between the two can be noticed at high concentration, where the value of the crystallinity for P$_{lin189}$ is higher than for P$_{lin40}$.
Since the two samples only differ in their molecular weight and thus in the concentration of chain-ends, which is higher for the smaller chains with respect to the bigger ones,  this result suggests a possible role of chain-ends in the reduction of water crystallization.

Moreover, comparing crystallinity at the same polymer concentration, it can be observed that the $\chi_c$ values in deuterium oxide are always larger  than the corresponding ones in water. This happens for linear chains and also for microgels, both non-deuterated and deuterated ones, indicating a higher affinity of the polymer, regardless of the structure, with water in the investigated low temperature/ high concentration range.
By comparing the effect of polymer deuteration on crystallization, it seems that h-P$_m$ have a larger suppressing effect than d-P$_m$, both in water and deuterium oxide, suggesting that the interactions of the non-deuterated microgels with solvents are higher and affect a larger number of solvent molecules as compared to deuterated microgels.

From these data, we can confirm that melting completely disappears for microgel samples at concentration above 60 wt\%. However, already at a concentration above 40 wt\%, we find that the value of $\chi_c$ is of the order of just a few percent. This may well explain the fact that in previous  EINS measurements~\cite{zanatta2018evidence}, the presence of D$_2$O crystals was already not detectable at 43 wt\% for microgels. The same considerations apply for linear chains, albeit at slightly larger concentration values~\cite{tavagnacco2021proteinlike}.

\subsection{Study of the glass transition}

\begin{figure}[hbtp]
\centering
\includegraphics[width=8.5cm]{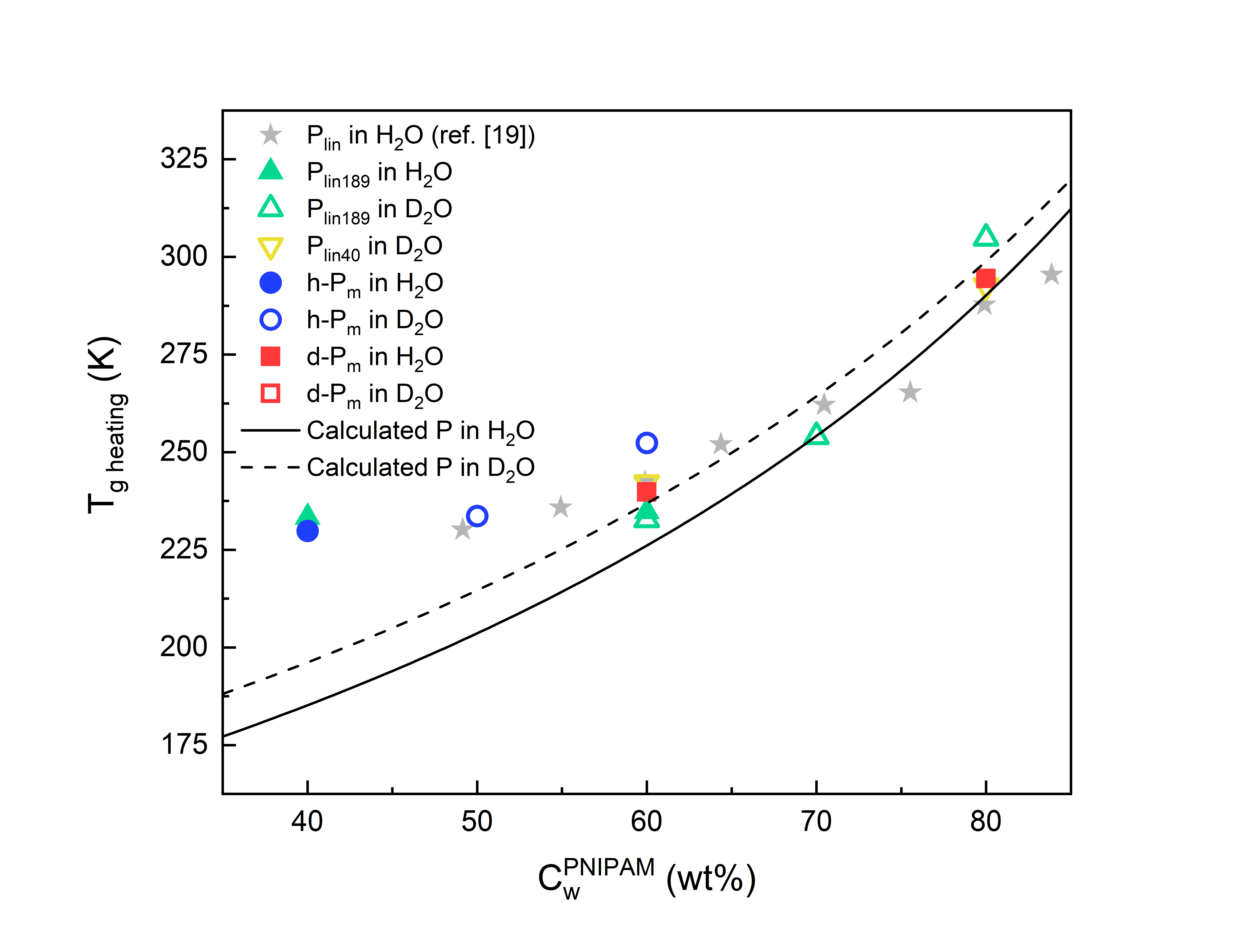}
\caption{Glass transition temperature, $T_g$, as a function of polymer concentration for each sample. Grey stars are data taken from~\cite{afroze2000phase}. The solid and dashed lines represent the concentration dependence of $T_g$ for PNIPAM in H$_2$O and D$_2$O, respectively, calculated using the Fox-equation~\cite{fox1956}: $1/T_g = w_1/T_{g1} + w_2/T_{g2}$, where $w_1$ and $T_{g1}$ are the weight fraction and glass transition temperature of pure PNIPAM, $w_2$ and $T_{g2}$ the weight fraction and glass transition temperature of pure solvent (either H$_2$O or D$_2$O) and $T_g$ the glass transition temperature of the dispersion. Errors on temperature values are estimated to be of the order of 0.5 K.}
\label{glasstransition}
\end{figure}

The glass transition of concentrated dispersions of PNIPAM linear chains and microgels was investigated by performing DSC measurements with cooling and heating steps at 5, 10 and 20 K/min. In Fig.~\ref{glasstransition} the glass transition temperature, $T_g$, measured in the heating step of the thermograms collected at 5 K/min, is reported for all samples where this was detected. For samples in water, the glass transition is observed only for concentrations equal and above 40 wt\%, while for samples in deuterium oxide this happens above 50 wt\%. However, we note that we could not find a clear evidence of the glass transition for all values of polymer concentrations measured between 40 and 80 wt\%. To this aim, we compared thermograms at different scanning rates and reported data points only for the conditions where this was found to be robust.  A reason why it appears to be difficult to locate the glass transition at intermediate concentrations is that this may be hidden by the concomitant melting or LCST/VPT peak. Fig.~\ref{glasstransition} also includes a comparison with previous results~\cite{afroze2000phase} and with the calculated concentration dependence of $T_g$ for miscible binary systems of PNIPAM in H$_2$O and D$_2$O, resulting by using the values of the glass transition temperatures for the pure solvent and PNIPAM. Overall, we find a good agreement between experimental points and such theoretical predictions at high concentrations, respectively, i.e., above 60 wt\%, while at lower concentrations there are deviations towards higher temperatures in the experimental data, as also found in Ref.~\cite{afroze2000phase}. Such deviation occurs when free water crystallization dominates the thermograms. Indeed, the mixing rule is based on the assumption that water is evenly distributed in the system. However, when a part of water crystallizes, such a condition is not valid. Therefore, the observed glass transition is the one of the amorphous phase whose composition levels off for C$_w$ below 40\%, when free water starts to crystallize.
In general, the difficulty to precisely detect a glass transition temperature in the whole investigated range of concentrations within our current analysis prevents us from assessing more quantitative statements about the dependence of $T_g$ on the specific polymer architecture and on the influence of isotopic substitution. It would be thus interesting to address this point in more detail in the future with more sensitive calorimetric protocols.

\subsection{Study of the Volume Phase transition / LCST}

\begin{figure}[hbtp]
\centering
\includegraphics[width=\textwidth]{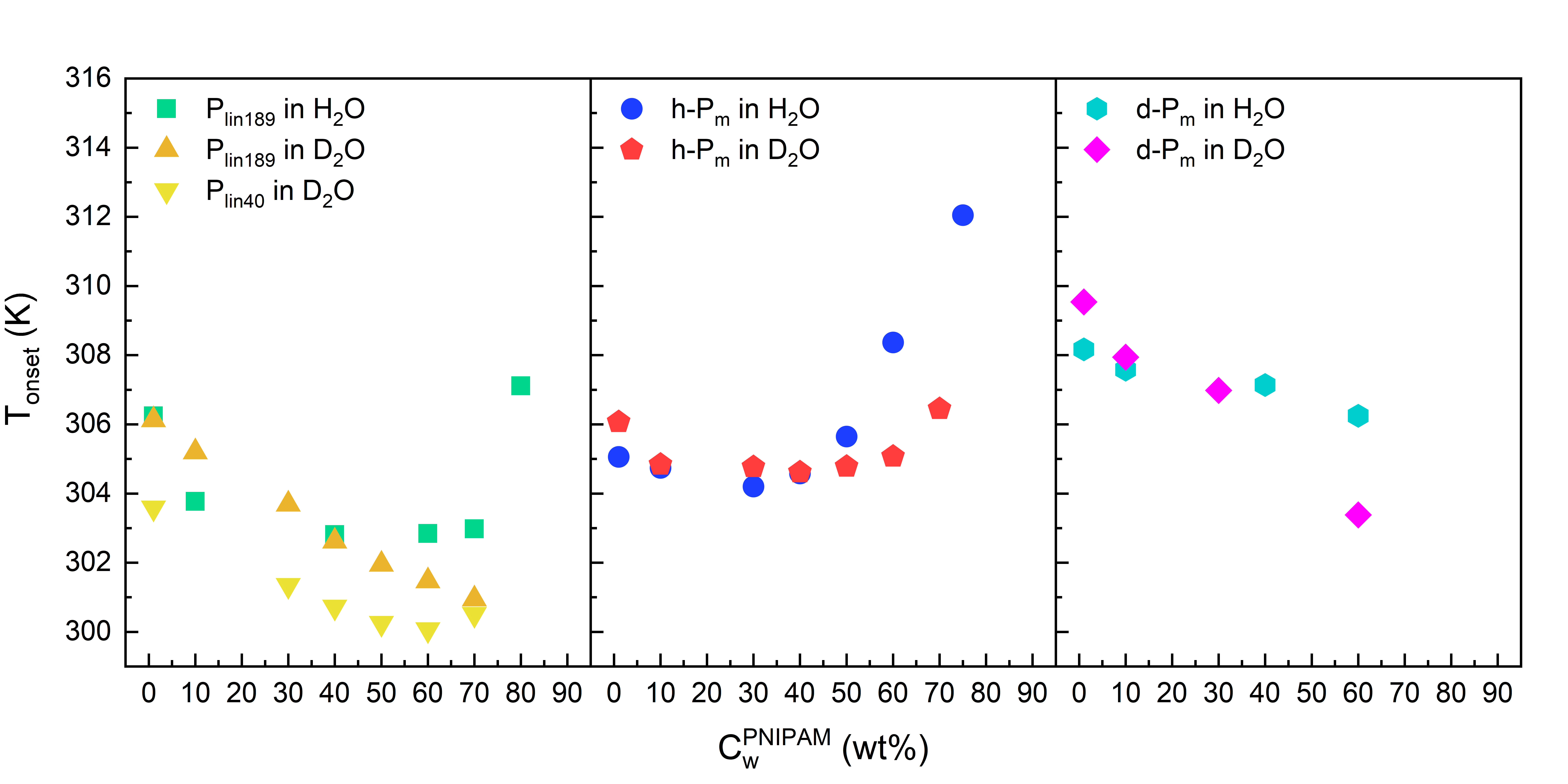}
\caption{Onset temperature $T_{onset}$ of the PNIPAM coil-to-globule (for P$_{lin189}$ and P$_{lin40}$) and volume phase (for h-P$_m$ and d-P$_m$) transition, as a function of polymer concentration, for all investigated samples. Errors on temperature values are estimated to be of the order of 0.5 K.}
\label{transition}
\end{figure}

After having examined the melting processes, we now focus on the behaviour of the PNIPAM coil-to-globule (for linear chains) and volume phase (for microgels) transition. In Fig. S3, an example of the evolution of the peak as a function of polymer concentration for the non-deuterated microgels in D$_2$O is reported. Some differences among the kind of structures investigated and their interaction with solvents can be observed as reported in Fig.~\ref{transition}, where the onset temperature, $T_{onset}$, which is the temperature at the beginning of the transition, is reported for all the samples as a function of polymer concentration.

The $T$-dependence is found to be similar for the LCST temperature of P$_{lin189}$ and for the VPT temperature of h-P$_m$ in H$_2$O: they are found to be just shifted at a lower value for linear chains with respect to microgels. However, in both cases we observe a minimum value of $T_{onset}$ occurring at around 40-50 wt\%, followed by an increase at higher concentration, in good agreement with previous works~\cite{afroze2000phase,zanatta2020atomic}. A different scenario occurs in D$_2$O: for microgels, the behaviour of $T_{onset}$ with concentration has the same trend as in H$_2$O, although at high concentration it does not rise as much as in water; for linear chains, instead, the minimum is almost not observable. Indeed, an almost linear decrease of T$_{onset}$ is evident for both P$_{lin189}$ and P$_{lin40}$ in D$_2$O up to concentration of 70 wt\%, above which the LCST is no longer detectable.
Turning to the deuterated microgels, it seems that  $T_{onset}$  only decreases with concentration, with a more pronounced reduction in D$_2$O than in H$_2$O, in analogy with what observed for h-P$_m$ and linear chains in the two different solvents.

Finally, comparing the behaviour between the two solvents, it appears that at low concentrations the transition temperature is always higher in deuterium oxide than in water, while increasing concentration above 40\% this trend is inverted, with values in D$_2$O always lower than those in H$_2$O.
The higher transition temperature detected for concentrated dispersions in H$_2$O indicates a greater affinity of PNIPAM for H$_2$O, as compared to D$_2$O, in the high concentration regime. A similar conclusion was drawn in a previous study of H$_2$O/D$_2$O exchange in dense and homogeneous PNIPAM thin films, whose hydration levels were below 35\% (v/v)~\cite{widmann2019hydration}.
In dilute regime an opposite trend of solvent affinity is found, with  affinity being higher in D$_2$O with respect to H$_2$O, in agreement with previous calorimetric experiments for PNIPAM chains at 0.1 wt\%~\cite{kujawa2001volumetric}. In addition, for very low concentrations (1-10 wt\%), we observe that d-P$_m$ microgels show values of $T_{onset}$ that are significantly higher ($\sim$ 308-310 K) than those of all other samples, both in H$_2$O and in D$_2$O.

For completeness, the enthalpies associated to the transition are reported as function of polymer concentration in Fig. S4. The values reach a maximum around 30-40 wt\% for linear chains and microgels. However, the enthalpy associated with the transition of the linear chains is slightly higher than the one of the microgels, as well as the one in D$_2$O as compared to the one in H$_2$O.

\subsection{Swelling behaviour of PNIPAM microgels}
\begin{figure}[hbtp]
\centering
\includegraphics[width=8.5cm]{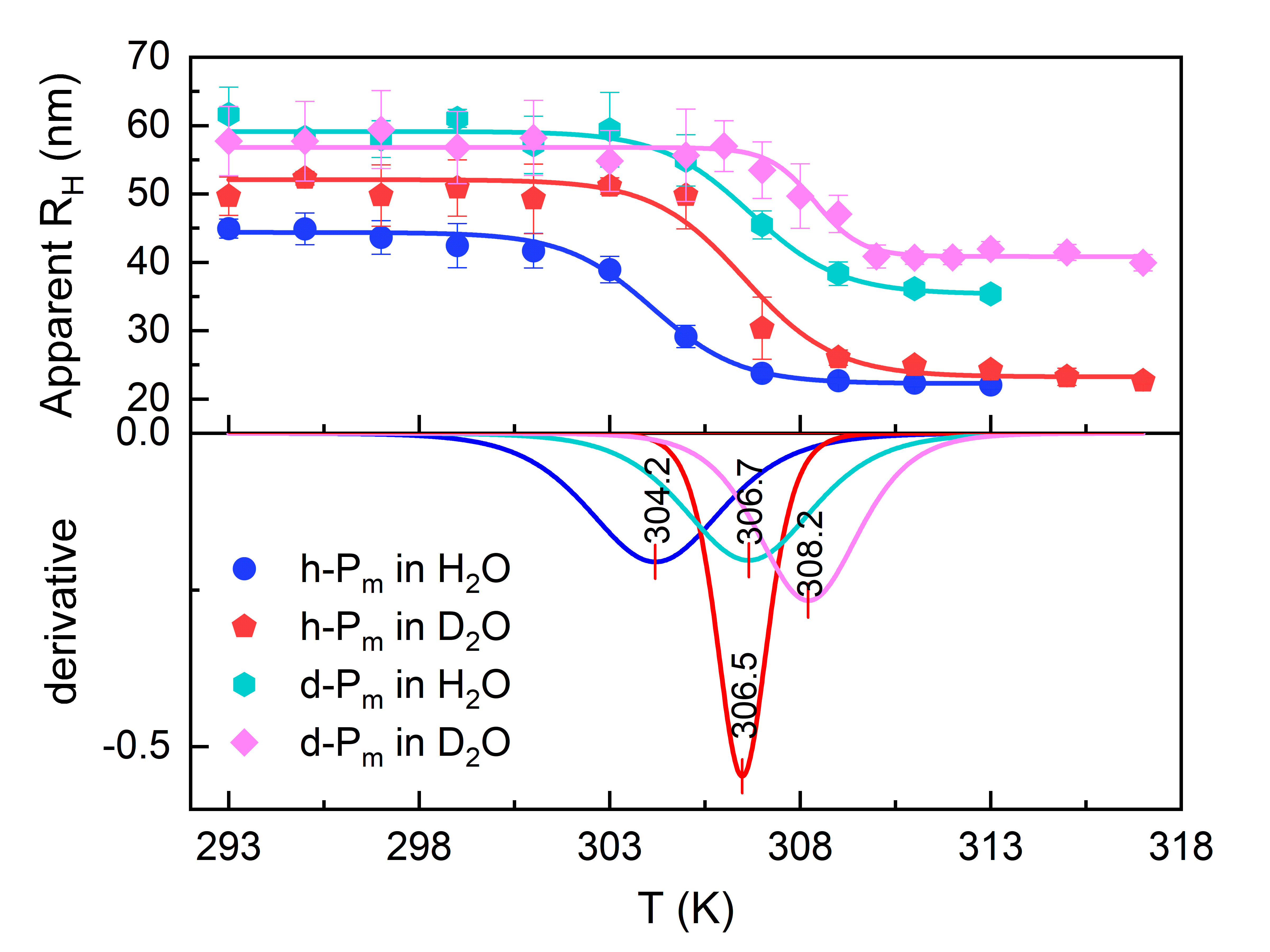}
\caption{(Top) Swelling curve of non-deuterated and deuterated PNIPAM microgels in H$_2$O and D$_2$O. (Bottom) Derivative of the fit of the experimental data in the upper part. The VPT temperatures of each sample are also reported. Data refer to microgel dispersions at a concentration of 0.01 wt\%.}
\label{DLS}
\end{figure}

In order to better quantify the differences induced by the specific solvent and by the nature of the microgels (deuterated vs non-deuterated), we perform dynamic light scattering measurements on the microgels in dilute conditions both in H$_2$O and D$_2$O.
 The corresponding swelling curves are reported in Fig.~\ref{DLS}
as a function of temperature. The experimental data were fitted with a Boltzmann equation~\cite{navarro2011modified, cors2019deuteration} and the derivatives of each function are shown in the bottom of the figure. The peaks of these derivatives represent the estimated VPT temperature at which microgels change from a highly hydrated swollen state into a partially dehydrated, collapsed state.

We find that the value of VPT temperature (VPTT) is considerably affected by the isotope substitution: in the case of h-P$_m$ in H$_2$O, the VPTT is observed at 304 K, around the classic value, while we find an increase of about two degrees both when deuterium is added to the polymer chains, i.e., d-P$_m$ in H$_2$O, or to the solvent, i.e., h-P$_m$ in D$_2$O. Namely, we find 306.5 and 306.7 K, respectively, for these two cases. These results are in good agreement with previous works on microgels in D$_2$O~\cite{nojd2018deswelling,cors2019deuteration}.
The higher VPTT value of h-P$_m$ in D$_2$O, with respect to H$_2$O, is consistent with the increased transition temperature detected for this system in the DSC experiments at the lowest polymer concentrations in D$_2$O, as compared to H$_2$O. A similar isotopic effect of the solvent was also reported for the VPTT of macroscopic PNIPAM hydrogels at low degree of crosslinking~\cite{shirota1998volume}, where the higher stability of the hydrogen-bonding interaction between the PNIPAM amide groups and the deuterium oxide molecules are ascribed to the lower vibrational energy of intermolecular hydrogen bond in D$_2$O with respect to H$_2$O due to the heavier mass~\cite{nemethy1964structure}.

The increase of LCST or VPTT in highly hydrated PNIPAM systems, detected moving from the protiated to deuterated isotopologues, can be instead ascribed to the lower lipophilicity of deuterated aliphatic groups as compared to the non-deuterated ones, because of the reduced vibrations of deuterons with respect to protons~\cite{turowski2003deuterium,wade1999deuterium}, which leads to a stabilization of the hydrated states. Similarly, the deuteration of non-exchangeable protons in globular proteins was found to reduce their denaturation temperature of few degrees~\cite{piszczek2011deuteration,white2018deuterium,nichols2020deuteration}, both in H$_2$O and D$_2$O~\cite{hattori1965effect}, confirming a deuteration-induced decrease in stability of globular, collapsed states.
Differently, for a few proteins in the native isotopic state, an increase of the denaturation temperature in D$_2$O was observed~\cite{hermans1959thermally,sportelli1985isotopic,makhatadze1995solvent,guzzi1998solvent}, which can be explained with the higher strength of inter-amide hydrogen bonds, when changing H with D atoms~\cite{calvin1959effect,li2021effect}. However, PNIPAM collapsed states still remain partially hydrated~\cite{pelton2010poly}, thus the contribution of inter-residue H bonds to its conformational transition is weak, becoming relevant only for highly concentrated dispersions, as also shown by the present calorimetric results. As a consequence, the prevailing factor in diluted aqueous environments moving from H$_2$O to D$_2$O is the reinforcement of polymer-solvent hydrogen bonds, leading to a stabilization of the more hydrated/swollen states.
When both the microgel and the solvent are deuterated, i.e., d-P$_m$ in D$_2$O, an additional shift of two degrees is observed, to even higher values (308.2 K). While there are a few studies that also report the shift for h-P$_m$ in D$_2$O~\cite{zanatta2020atomic,nigro2017dynamical,berndt2006temperature,niebuur2018formation,niebuur2019kinetics} and for d-P$_m$ in H$_2$O~\cite{sbeih2019structural,brugnoni2019synthesis,widmann2019hydration,crowther1999poly}, we are not aware of any previous report of this effect.
A tentative explanation of this intriguing, new result is that the combination of the two effects discussed above sums up, giving rise to a ``double'' shift of the transition temperature. It is also worth noting that the transition is sharper in D$_2$O than in H$_2$O for both h-P$_m$ and d-P$_m$, as can be seen from the peak broadness of the derivative, that is especially narrow for the non-deuterated polymer in D$_2$O. This is in good agreement with other works on PNIPAM-based materials~\cite{nigro2017swelling,nigro2021chemical,shirota1998volume,wang1999light}. The shape of the transition for h-P$_m$ and d-P$_m$ in H$_2$O is instead almost identical.

Finally, a comment is due on the swelling degree, $\alpha$, that is defined as the ratio between the particle radius in the swollen state (at 293 K) and that in the collapsed state (at 313 or 317 K). This quantity also seems to be affected by either the isotope or the solvent substitution. In fact, while h-P$_m$ microgels have a comparable size both in H$_2$O and D$_2$O in the collapsed form, this does not apply to the swollen one, leading to a change in the swelling degree from roughly 2 to 2.5, respectively. A different  scenario occurs for d-P$_m$, where the collapsed state is larger in D$_2$O, while the swollen ones are similar. This leads to $\alpha$ roughly equal to 1.7 in H$_2$O and 1.4 in D$_2$O. However, these values are significantly lower than the ones of h-P$_m$ in either H$_2$O or D$_2$O. Altogether these features suggest a lower energy of the polymer-polymer interactions in the case of d-P$_m$ with respect to h-P$_m$ and even a lower difference (polymer-polymer, polymer-solvent) in the case of the deuterated microgels.

\section{Conclusions}

In this work, we investigated how the polymer architecture influences water behaviour in concentrated samples of poly(\emph{N}-isopropylacrylamide): two PNIPAM linear polymers at two different molecular weights and two weakly crosslinked microgels, one of which synthesised from a deuterated monomer. The aim was to provide a detailed investigation of water crystallization in poorly hydrated PNIPAM suspensions and in particular to study the possible effects of isotope substitution.
Indeed, recent elastic neutron scattering measurements pointed out that hydrated PNIPAM powders undergo a protein-like dynamical transition~\cite{tavagnacco2019water,zanatta2018evidence} and suggest that hydration water does not crystallize even at polymer concentrations as low as 50 wt\%. Since the protein-like dynamical transition is often considered to be driven by the atomic dynamics of the solvent~\cite{schiro2015translational,bellissent2016water}, independent and reliable information about the amorphous or crystalline state of hydration water is extremely valuable for a robust interpretation of neutron data. Moreover, the subject of avoided crystallization in confined water is per se a long standing problem in the physics of water itself~\cite{gallo2021advances,debenedetti2020second}. In addition, since the previous works were mainly based on neutron scattering experiments in D$_2$O environment, we also compare the behaviour of PNIPAM samples in H$_2$O and D$_2$O, in order to highlight possible differences between the two solvents. Through calorimetric analysis in a wide range of temperatures and polymer concentrations, we monitored, in particular, the behaviour of the melting peak and of the one corresponding to the PNIPAM thermoresponsive transition (LCST and VPT for linear chains and microgels, respectively). We also, when possible, paid attention to the occurrence of a glass transition in the samples.

Overall, we found strong indications that PNIPAM microgels are more efficient in suppressing water crystallization with respect to linear chains, since they prevent water crystallization at lower polymer concentration (60\% against 75\%, respectively). Moreover, in this high-concentration regime (50-80 wt\%), we observed that solvent-polymer interactions are preferred to those between solvent molecules to a larger extent when the solvent is H$_2$O, as shown by the higher degree of crystallinity obtained in D$_2$O than in H$_2$O.
This greater affinity with aqueous solvent in concentrated dispersions is also shown in the PNIPAM conformational transition: indeed, the transition temperature above $\geq40$ wt\% polymer concentration is found to be higher for samples in H$_2$O than in D$_2$O, indicating that the polymer finds water a ``better'' solvent than deuterium oxide, as proposed in~\cite{widmann2019hydration}.
This may ascribed to the fact that, at high polymer concentrations, the contribution of polymer-polymer H-bonds in the transition become more important.
In D$_2$O the amides are ND(CO), so that the inter-residue H-bonds are stronger than in H$_2$O, where they occur between NH(CO) groups. Thus, it is more convenient to increase the number of polymer-polymer H-bonds in D$_2$O than in H$_2$O, leading to the anticipation of the transition.
Instead, in the low-concentration regime, this aspect is reversed, since the transition happens at lower temperature in H$_2$O, confirming that the polymer-D$_2$O interactions are stronger than the polymer-H$_2$O interactions in this regime, as also supported by the swelling behaviour and DLS measurements of the microgels in very dilute conditions.

Finally, we analysed the isotope substitution not only by changing the solvent, but also investigating a
deuterated microgel. We found that, in the low-concentration regime, the use of either deuterated particles in H$_2$O or non-deuterated microgels in D$_2$O shifts the VPT temperature by roughly two degrees, as previously observed in several works~\cite{zanatta2020atomic,nigro2017dynamical,berndt2006temperature,sbeih2019structural}, while the consideration of deuterated particles in D$_2$O induces an unexpected ``double'' shift of the VPT temperature, occurring roughly four degrees later than the standard one. The increase of the VPTT from h-P$_m$ to d-P$_m$ in D$_2$O remains to be clarified. We tentatively attribute this feature to the effect of PNIPAM hydrophobic groups. In particular, the protiated hydrophobic groups are more lipophilic (and therefore less affine to water, be it H$_2$O or D$_2$O) than the deuterated ones. Thus, the increase of contacts between hydrophobic groups occurring at the VPT is enthalpically more advantageous for h-P$_m$ than for d-P$_m$, even in D$_2$O, as it is in H$_2$O, leading to a lower VPTT. Overall the double VPTT-increasing effect observed for non-deuterated microgels in D$_2$O may thus be due to two separate contributions, arising from different parts of the polymer: (i) an effect arising from the hydrophilic groups (amides) and (ii) an effect arising from hydrophobic groups. In the first case, going from H$_2$O to D$_2$O, the amide-water hydrogen bonds are strengthened, so that a loss of water is more disadvantageous in D$_2$O, which increases the transition $\Delta H$ and thus the VPTT, with the phenomenon being identical for h-P$_m$ and d-P$_m$. In the second case, going from h-P$_m$ to d-P$_m$, the interactions between aliphatic groups are weakened, so that so that the VPT, leading to an increase of such interactions, is less advantageous for d-P$_m$ than for h-P$_m$ (regardless of whether this happens in H$_2$O or D$_2$O). These factors may thus explain the additivity of the ``double'' shift of the VPT temperature.

We believe that our results will be valuable to correctly interpret past and future neutron scattering measurements of these highly concentrated samples, particularly to address the role of solvent crystallization and to differentiate the effects due to the use of D$_2$O vs H$_2$O. Given the relevance of the low-temperature dynamical transition in biological samples and the shortage of similar calorimetric studies for these systems~\cite{kyritsis2012water}, our work could be important to address these issues also for protein dispersions, shedding further light on the role of the solvent in this debated topic~\cite{tavagnacco2019water}.

\section{Acknowledgement}

We acknowledge financial support from MIUR (FARE project R16XLE2X3L, SOFTART), from the European Research Council (ERC Consolidator Grant 681597, MIMIC) and from Regione Lazio (through L.R. 13/08 Progetto
Gruppo di Ricerca MICROARTE n. prot. A0375-2020-36515).

\end{document}